\documentstyle[aps,preprint]{revtex}

\begin{document}
\newcommand{\be}{\begin{equation}}
\newcommand{\ee}{\end{equation}}
\newcommand{\bea}{\begin{eqnarray}}
\newcommand{\eea}{\end{eqnarray}}
\newcommand{\eps}{\varepsilon}
\begin{center} 
{\bf { \large  Non-adiabatic molecular Hamiltonian. }}
\end{center}
\begin{center}
{\large Canonical transformation coupling electronic }
\end{center}
\begin{center}
{\large and vibrational motions.}
\end{center}  
{\it
\begin{center}
Ivan Huba\v{c}
$^{a,b}\footnote{Permanent and correspondence address}^{,c}$, Peter Babinec  $^{a,b}$, Martin
 Pol\'{a}\v{s}ek \ $^{c}$, 
J\'{a}n Urban  $^{b}$, \\ Pavel Mach  $^{b}$ 
, Jozef M\'{a}\v{s}ik  $^{b}$ and  Jerzy Leszczy\'{n}ski  $^{a}$
\end{center}
\begin{center} 
$^a$Department of Chemistry, Jackson State University, 1400 Lynch Street, \\
P.O. Box 17910, Jackson, MS 39217, USA\\ 
$^b$Division of Chemical Physics, Faculty of Mathematics
and Physics, Comenius University, \\Mlynsk\'{a} dolina F1, 842 15 Bratislava,
Slovakia \\
$^c$ Institute of Physics, Faculty of Science, Silesian University,
 Bezru\v{c}ovo n\'{a}m. 13,\\
746 01 Opava, Czech Republic
\end{center}
}
{\bf Abstract}

The coupling of electronic and vibrational motion is studied
by two canonical transformations namely normal coordinate
transformation and momentum transformation on molecular Hamiltonian.
It is shown that by these transformations we can pass from crude 
approximation to adiabatic approximation and then to non-adiabatic
(diabatic) Hamiltonian. This leads to renormalized fermions and 
renormalized diabatic
phonons. Simple calculations on $H_{2}$,  $HD$, and $D_{2}$ systems are 
performed and compared with previous approaches. Problem of
electronic quasi-degeneracy is discussed.
\newpage

\section{Introduction}

Many atomic systems (e.g. molecules and/or crystals) are systems 
with interacting electrons and nuclei and can be thus described by   
Schr\"odinger equation
\begin{equation}
\label{e1}
	H \Psi\ = \ E \Psi
\end{equation}
In general case the number of degrees of freedom of such systems with
coulomb interaction is too large and even using high-performance
computers to solve this equation becomes impossible.
Therefore the only possibility to solve eq. (\ref{e1}) is to suggest 
some approximations \cite{Born-Huang,longuet,Koppel,Wilson,Sutcliffe,yarkony}. The most important approximation and also most 
often used is the Born-Oppenheimer (BO)\cite{Born27} and the adiabatic approximation.
This approximation is based on the fact that masses of nuclei are  
1870 times heavier than the masses of electrons. This leads to the idea 
of potential energy surface. Beside the many attemps to go 
beyond the BO approximation and many different approaches
cartain facts are not clear completely. In this paper we decided
to study the coupling of electronic and vibrational motions by two 
canonical transformations namely normal coordinate and momentum
transformations. Our approach is similar to quasiparticle
transformations often done in solid state physics. In order to make
our approach more transparent we repeat here main features
of adiabatic approximation. We follow the arguments of recent
Kutzelnigg's paper \cite{Kutzi1}.\\
Let us write the total molecular Hamiltonian as  
\begin{equation} \label{e2}
	H \ = T_{N}(R) + E_{NN}(R) + H_{EN}(r,R) +  H_{EE}(r)
\end{equation}
where  $T_{N}(R)$ is the kinetic energy of nuclei, 
\ $E_{NN}(R)$\
is the interaction between nuclei,  $R$  denotes nuclear coordinates,
r denotes electronic coordinates, and   $H_{EN}(r,R) +  H_{EE}(r)$ 
is the electronic Hamiltonian  

\begin{equation} \label{e3}
H_{EN}(r,R) +  H_{EE}(r) \ = T_{E}(r) + U_{EN}(r,R) + H_{EE}(r)
\end{equation}
where $T_{N}(r)$\ is the kinetic energy of electrons,
 \ $U_{EN}(r,R)$ \ is the electron--nuclei interaction term and 
\ $H_{EE}(r)$\ represents the electron-electron interaction.
In Born and Huang \cite{Born-Huang,Davidov} approach the total wavefunction depending on 
the nuclear coordinates \ $R$ \ and the electronic  coordinates \ $r$ \
is expanded as  
\begin{equation}  \label{e4}
\Psi(r,R) \ = \sum_{k} \psi_{k}(r,R)\chi_{k} (R),
\end{equation}
where \ $\psi_{k}$ \ are a complete set of known functions of 
 \ $r$ \ that depend parametrically on the  nuclear coordinates \ $R$ \
and where  the  $\chi_{k}(R)$  are regarded as unknown. The \ $\psi_{k}$ \
are conventionally chosen as a set of eigenfunctions of the clamped nuclei 
(CN) Hamiltonian, but this is not necessarily a good choice, since this is 
usually not complete (without, the corresponding continuum functions,
which one cannot include anyway). Both the BO and adiabatic
approximation can be based on choosing a single term in eq. (\ref{e4})

\begin{equation}  \label{e5}
\Psi(r,R)  = \psi_{k}(r,R)\chi_{k} (R),
\end{equation}
This is referred as BO ansatz. This ansatz is taken as a variational
trial function.
Terms beyond the leading order in   $m/M$  are neglected (   $m$ 
is the electronic and   $M$  is nuclear mass, respectively). 
The problem with expansion  (\ref{e4}) is that functions
  $\psi(r,R)$  contain except bound states also continuum function 
since it includes the centre of mass (COM) motion. Variation principle
does not apply to continuum states. To avoid this problem we can 
separate COM motion. The remaining Hamiltonian for the relative motion
of nuclei and electrons has then bound state solution. But there is
a problem, because this separation mixes electronic with vibrational
coordinates and also there is a question how to define molecule-fixed 
coordinate system. This is in detail discussed  by Sutcliffe \cite{Sutcliffe} 
. In the recent paper by Kutzelnigg \cite{Kutzi1} this problem is
also discussed and it is shown how to derive in a rigorous manner
adiabatic corrections using so called Born--Handy ansatz. There are few 
important steps to arrive at formula for a diabatic
corrections. Firstly, one separates off COM motion. Secondly,
(very important step) one does not specify the relative
coordinates (which are to some extent arbitrary). In this way
one arrives at relative Hamiltonian $H_{rel}$ \cite{Kutzi1} with
trial wavefunction $\Psi_{rel}$. If we make BO ansatz
\begin{equation}  \label{A}
\Psi_{rel}  = \psi(\cdots \rho_{lk},\cdots \rho_{\mu \nu}\cdots)
\chi(\cdots \rho_{\mu \nu}\cdots) 
\end{equation}
where $\rho_{lk}$, $\rho_{\mu \nu}$ are non-specified relative
coordinates and $\psi$ is chosen as a solution of the CN Schr\"{o}dinger
equation. The adiabatic correction $\Delta E$ take very simple form

\begin{equation} \label{B}
\Delta E = - \frac{1}{2} < \Psi | \sum_{ \mu} M_{\mu}\ \nabla_{\mu}^{2} | \Psi >
\end{equation}

This formula was used by Handy \cite{Handy}. It was used previously also by
Sellers and Pulay \cite{Sellers}. 
 (See also
 Davidov \cite{Davidov} for derivation). For practical calculation
the identity

\begin{equation} \label{C}
\int \psi \frac{\partial^{2}}{\partial Q_{i}^{2}} \psi\ dr_{i}=
-\int \left( \frac{\partial \psi}{\partial Q_{i}} \right )^{2} dr_{i}
\end{equation}
can be used.\\
Note that practically in any textbook \cite{Davidov} the validity
of BO approximation is justified only when

\begin{equation} \label{BO}
\frac{\hbar \omega}{|E_{n}-E_{m}|} \ll 1
\end{equation}
when $\omega$ is the frequency of harmonic vibrations arround the point 
$R_{0}$. \\
The aim of this paper is twofold: \\
i, We show how starting with
molecular Hamiltonian (\ref{e2}) in crude adiabatic representation
we arrive at adiabatic Hamiltonian by performing canonical
transformations which mix together the electronic and vibrational
motions (through normal coordinates). We derive simple formulae
for adiabatic corrections, similar to eq. (\ref{B}). \\
ii,  We generalize canonical transformations (through
momenta) arriving at non--adiabatic Hamiltonian. We introduce
the idea of quasiparticles (renormalized electrons) and present the formulae how to
obtain the "orbital energies", "correlation corrections"
and non--adiabatic frequencies for these  quasiparticles. Finally,
we perform some simple model calculations to demonstrate how
the method works.

\section{Theory}

Let us start with electronic Hamiltonian (\ref{e3}) which we denote

\begin{equation} \label{e6}
H_{EN}(r,R) + H_{EE}(r) \ = \ h + \nu^{0} ,
\end{equation}
where  $h$  is the one-electron part representing the kinetic energy of the
electrons and electron-nuclear attraction term, and  $v^{0}$  is the two
electron part of the Hamiltonian corresponding to electron-electron repulsion
term. For the purpose of
diagrammatic many-body perturbation theory it will be efficient to work in
second quantization formalism. The electronic Hamiltonian
(\ref{e6}) has the form
\begin{eqnarray} \label{e7}
H_{EN} + H_{EE} \ &=& \ \sum_{PQ} < P | h | Q > a_{P}^{+} a_{Q}  \nonumber 
\\[2mm]
&+& \ \frac{1}{2} \sum_{PQRS} < PQ |v^{0}| RS > 
a_{P}^{+} a_{Q}^{+} a_{S} a_{R}  
\end{eqnarray}
where  $a_{P}^{+}(a_{Q})$  is the creation (annihilation) operator for
electrons in the spinorbital basis  $|P>, |Q>, \cdots$  . If we apply
 the Wick theorem to (\ref{e7}) we can write this equation as
\vspace{1cm}
\begin{eqnarray} \label{e8}
H_{EN} + H_{EE} & = & \ \sum_{I}h_{II} + \frac{1}{2} \sum_{IJ} \left(v_{IJIJ}^{0} -
v_{IJJI}^{0} \right) \ + \nonumber \\[2mm]
&+&  \sum_{PQ}h_{PQ}N \left[ a_{P}^{+}a_{Q} \right] + \sum_{PQI} \left(
v_{PIQI}^{0} - v_{PIIQ}
^{0} \right) N \left[ a_{P}^{+} a_{Q} \right] \nonumber  \\[2mm]
&+& \ \frac{1}{2} \sum_{PQRS} v_{PQRS}^{0} N \left[ a_{P}^{+} a_{Q}^{+} a_{S}
a_{R} \right] 
\end{eqnarray}
where  $v_{ABAB}^{0}(v_{ABBA}^{0})$  denotes the coulomb (exchange) integral.
One possibility is to work the within crude representation in which
the spinorbital basis  $|P>, |Q>, \cdots$  is determined at some fixed
(equilibrium coordinate  $R_{o}$). Note that Hamiltonian (\ref{e8}) has
$3N-6$ degrees of freedom (in fact 3N degrees of which 6 are zero).
Hamiltonian (\ref{e8}) has only bound-state solutions. Let us divide individual
terms of the Hamiltonian (\ref{e8}) into two parts. Namely caculated at point
 $R_{o}$ and the terms which
are shifted with respect to term at \ $R_{o}$ (we use prime to denote
these terms). The electronic Hamiltonian (\ref{e8}) can be rewritten as

\begin{eqnarray} \label{e9}
H_{EN} + H_{EE} & = & \ E_{SCF}^{0} + h_{SCF}^{'} + \sum_{P} \varepsilon_{P}N
\left[ a_{P}^{+} a_{P} \right]  \nonumber \\[2mm]
&+& \sum_{PQ} h_{PQ}^{'} N \left[ a_{P}^{+} a_{Q} \right] \nonumber \\[2mm] 
&+& \frac{1}{2} \sum_{PQRS} v_{PQRS}^{0}
N \left[ a_{P}^{+} a_{Q}^{+} a_{S} a_{R} \right] 
\end{eqnarray}
where  $E_{SCF}^{0}$  is the Hartree-Fock energy calculated at the point
 $R_{0}$  , and  $h_{SCF}^{'}$  is the shift in the Hartree-Fock energy with
respect to the point other than  $R_{0}$. The same is true for one-particle
operator of (\ref{e9}), where  $\varepsilon_{P}$  are the one-particle Hartree-Fock
energies calculated at point  $R_{0}$. 
The correlation operator is not changed
because it does not depend on nuclear coordinates  $R$ . For the notation
see \cite{IJQC}.
Let us perform the Taylor expansion for the energies  $E_{NN}$  and
 $u_{SCF}$  around the point  $R_{0}.$ 
\begin{equation} \label{e10}
E_{NN} \ = \ E_{NN}^{(0)} + E_{NN}^{'} \ = \ \sum_{i=0}^{\infty} E_{NN}^{(i)}
\end{equation}
and
\begin{equation} \label{e11}
u_{SCF} \ = \ u_{SCF}^{(0)} + u_{SCF}^{'} \ = \ \sum_{i=0}^{\infty} u_{SCF}^{(i)}
\end{equation}
Using (\ref{e10}) and (\ref{e11}) we can rewrite our Hamiltonian (\ref{e9})
 in the form
\begin{eqnarray} \label{e12}
H & =& \ E_{NN}^{(0)} + E_{SCF}^{(0)} + \sum_{r} \hbar \omega_{r} \left( b_{r}^{+}b_{r}
 + \frac{1}{2} \right) + u_{SCF}^{(2)}
 + \sum
_{P} \varepsilon_{P} N \left[ a_{P}^{+} a_{P} \right] \nonumber \\[2mm]
&+&  \frac{1}{2} \sum_{PQRS} v_{PQRS}^{0} N \left[ a_{P}^{+} a_{Q}^{+} a_{S}
a_{R} \right] \nonumber \\[2mm] 
&+& \ E
_{NN}^{'} - E_{NN}^{(2)}  
+ \ u_{SCF}^{'} \ - \ u_{SCF}^{(2)} \ + \ \sum_{PQ} u_{PQ}^{'} N \left[ a_{P}^
{+} a_{Q} \right] 
\end{eqnarray}
where
\begin{equation} \label{e13}
 \sum_{r} \hbar \omega_{r} \left( b_{r}^{+}b_{r} + \frac{1}{2} \right)
=T_{N} + E_{NN}^{(2)} + u_{SCF}^{(2)} \ 
\end{equation}
$\omega_{r}$  is the frequency of the harmonic oscilator and 
 $b^{+}$  ($b$) are boson (phonon) creation (annihilation) operators.
In order to use the perturbation theory we have to split the Hamiltonian (\ref{e12}) onto
the unperturbed part  $H_{0}$ and the perturbation  $H^{'}$ 
\begin{equation} \label{e14}
H \ = \ H_{0} + H^{'}
\end{equation}

Due to the crude approximation, we can partition the Hamiltonian 
(\ref{e12}) in the following way
\begin{equation} \label{e15}
H_{0} \ = \ E_{NN}^{(0)} + E_{SCF}^{(0)} + \sum_{P} \varepsilon_{P} N \left[
a_{P}^{+} a_{P} \right] + \sum
_{r} \hbar \omega_{r} \left( b_{r}^{+} b_{r} + \frac{1}{2} \right)
\end{equation}
and
\begin{equation} \label{e16}
H^{'} \ = \ H_{E}^{'} + H_{F}^{'} + H_{I}^{'} .
\end{equation}
Where  $H^{'}$  contains all the  terms in (\ref{e12}) except (\ref{e15}).
In eq. (\ref{e12}) all quantities were defined through the cartesian coordinates.
For further purposes it will be natural to work in normal coordinates \ $\{ B_{r} \}.$ \
The normal coordinate in second quantized formalism is given as
\begin{equation} \label{e17}
B_{r} = b_{r} + b_{r}^{+}
\end{equation}
If we transform Hamiltonian (\ref{e12}) into normal coordinates we arrive at the
following expresions \cite{IJQC}

\begin{eqnarray} \label{e18}
H & =&  E_{NN}^{(0)} + E_{SCF}^{(0)} + \sum_{P} \varepsilon _{P} N \left[
a_{P}^{+}
a_{P} \right] + \sum_{r} \hbar \omega_{r} \left( b_{r}^{+} b_{r} 
+ \frac{1}{2}\right)  \nonumber \\[2mm]
&+&  \frac{1}{2} \sum_{PQRS} v_{PQRS}^{0} N
\left[ a_{P}^{+}a_{Q}^{+}a_{S} a_{R} \right] \ \ \left\{ \ \equiv H_{E}^{'}  \right\}
  \nonumber \\[2mm]
&+&   \sum_{n = 1, (n \neq 2)}^{\infty} \sum_{k=0}^{[n/2]} 
\left( {\bf E}_{NN}^{(k,n - 2k)} + {\bf u}_{SCF}^{(k,n - 2k)} \right).  
{\bf B}^{(n - 2k)} \ \ \left\{ \ \equiv H_{F}^{'} \right\}    \nonumber \\[2mm]
&+& \sum_{n=1}^{\infty} \sum_{k=0}^{[n/2]} \sum_{PQ}
{\bf u}_{PQ}^{(k,n-2k)}  . {\bf B}^{(n-2k)} N[a_{P}^{+} a_{Q}]
\ \ \left\{ \ \equiv H_{I}^{'} \right\}  
\end{eqnarray}
The term  $H_{E}^{'}$  is the electron correlation operator, the term
 $H_{F}^{'}$  corresponds to phonon-phonon interaction and  $H_{I}^{'}$ 
corresponds to electron-phonon interaction.
If we analyze the last term  $H_{I}^{'}$  we see that when using crude
approximation this corresponds to such phonons that force constant
in eq. (\ref{e13}) is given as a second derivative of electron--nuclei
 interaction with respect to normal coordinates. 
Because we used crude adiabatic approximation in which minimum of the energy
is at the point  $R_{o}$  , this is also reflected by basis set
used. Therefore this approximation does not properly describes the physical
vibrations i.e. if we move the nuclei, electrons are distributed according 
to the minimum of energy at point  $R_{o}$  and they do not feel
correspondingly the  $R$  dependence. The perturbation term   $H_{I}^{'}$ 
which corresponds to electron--phonon interaction is too large
and thus perturbation theory based on splitting given by eq. (\ref{e15}, 
\ref{e16}) will not
converge \cite{unpublished}. Natural way to improve this situation will be
 to use
basis set which is generally   $R$ dependent. We can do this in 
second--quantized formalism in a way that we pass from electron
creation (annihilation) operators  $a_{P}^{+}$ ($a_{Q}$)
which act on   $R_{o}$  dependent basis set to a new fermion
creation (annihilation) operators   $\stackrel{-}{a}_{P}^{+}$ 
 ( $\stackrel{-}{a}_{Q}$) which act on R dependent basis. 
Similar transformation was studied for solid state theory by
Wagner \cite{wag1}, who also discuss the convergency properties
of adiabatic approximation \cite{wag2}.
This we can achieve by
canonical transformation passing from old electron operators 
 $a_{P}^{+}$\ ($a_{Q}$) to new operators  $\stackrel{-}{a}_{P}^{+}$ \ 
($\stackrel{-}{a}_{P}$) through normal coordinates 
\ $B_{r}$ . In this way we can pass from crude adiabatic Hamiltonian to 
what is called clamped nucleus Hamiltonian and corresponding 
clamped nucleus wavefunction  $\Psi(r,R)$. The proof that this is a
canonical transformation is in \cite{IJQC}.

\begin{equation} \label{e19}
\stackrel{-}{a}_{P}  =  a_{P} + \sum_{Q} \sum_{k=1}^{\infty} \frac{1}{k!}
\sum_{r_{1} \ldots r_{k}} C_{PQ}^{r_{1} \ldots r_{k}}  B_{r_{1}} \ldots
B_{r_{k}} \ a_{Q} \ 
\end{equation}

\begin{equation} \label{e20}
\stackrel{-}{a}_{P}^{+} \ = \ a_{P}^{+} + \sum_{Q} \sum_{k=1}^{\infty} \frac{1}{k!}
\sum_{r_{1} \ldots r_{k}} C_{PQ}^{r_{1} \ldots r_{k}} \ B_{r_{1}} \ldots
B_{r_{k}} \ a_{Q}^{+} \ ,
\end{equation}
where $B_{r}$ are second quantized normal coordinates. \\
In short notation we can also write \cite{IJQC}
\begin{eqnarray} \label{e21}
\stackrel{-}{a}_{P} & =&  \sum_{Q} \sum_{k=0}^{\infty} \frac{1}{k!} {\bf C}_{PQ}
^{(k)} \ {\bf B}^{k} a_{Q} \nonumber \\
& =& \sum_{Q} \sum_{k=0}^{\infty} C_{PQ}^{(k)} a_{Q} \
   \nonumber \\[2mm]
&=&  \sum_{Q} C_{PQ} a_{Q} 
\end{eqnarray}
\begin{eqnarray} \label{e22}
\stackrel{-}{a}_{P}^{+} & =& \sum_{Q} \sum_{k=0}^{\infty} \frac{1}{k!} {\bf C}_{PQ}
^{(k)*} \ . \ {\bf B}^{k} a_{Q}^{+} \nonumber \\[2mm]
 &=& \sum_{Q} \sum_{k=0}^{\infty}
C_{PQ}^{(k)+} a_{Q}^{+}   \nonumber \\[2mm]
&=& \ \sum_{Q} \ C_{PQ}^{+} a_{Q}^{+} 
\end{eqnarray}

We also perform analogous canonical transformation for phonons

\begin{equation} \label{e23}
\stackrel{-}{b}_{r} \ = b_{r} + \sum_{PQ} \sum_{k=0}^{\infty} \frac{1}{k!} \sum
_{s_{1} \ldots s_{k}} d_{rPQ}^{s_{1} \ldots s_{k}} B_{s_{1}} \ldots B_{s_{k}}
a_{P}^{+} a_{Q} \ ,
\end{equation}

\begin{equation} \label{e24}
\stackrel{-}{b}^{+}_{r} \ = \ b_{r}^{+} \ + \ \sum_{PQ} \sum_{k=0}^{\infty}
\frac{1}{k!} \sum_{s_{1} \ldots s_{k}} d_{rPQ}^{s_{1} \ldots s_{r}^{*}}
B_{s_{1}} \ldots B_{s_{k}} a_{Q}^{+} a_{P}
\end{equation}
The coefficients \ $C_{PQ}$ \ ($C_{PQ}^{+}$) \ in eqs. (\ref{e21}, \ref{e22}) 
are determined so that  $\stackrel{-}{a}_{P}$ 
 ($\stackrel{-}{a}_{P}^{+}$  ) satisfy
 fermion anticommutation relation.
The coefficients  $d_{rPQ}$ (  $d_{rPQ}^{+}$  ) in eqs. (\ref{e23}, \ref{e24})
are determined so that  $\stackrel{-}{b}_{r}$ \ ($\stackrel{-}{b}_{r}^{+}$\ ) satisfy boson 
commutation relation. Finally we ask fermions \ $\stackrel{-}{a}_{P}$ 
 ($\stackrel{-}{a}_{P}^{+}$  )
to commute with bosons  $\stackrel{-}{b}_{r}$ \ ($\stackrel{-}{b}_{r}^{+}$\ ). 
This means that we can 
write similarly as in (\ref{e5}) the total wave function  $\Psi(r,R)$ 
as a product of fermion wave function  $\psi_{k}(r,R)$ and boson 
wave function  $\chi_{k}$  as $\Phi(r,R)$

\begin{equation} \label{e25}
\Phi(r,R)  = \psi_{k}(r,R)\chi_{k} (R),
\end{equation}
It is easy to show that we have two invariants of transformations,
namely number operator of fermions

\begin{equation} \label{e26}
\stackrel{-}{N} \ = \ N \ .
\end{equation}
and normal coordinate
\begin{equation} \label{e27}
\stackrel{-}{\bf B} \ = \ {\bf B}
\end{equation}
The next step is that we find inverse transormations to (\ref{e21}-\ref{e24})
and substitute these inverse transormations into eq. (\ref{e18}) and then applying Wick
theorem, we requantize the whole Hamiltonian (\ref{e12}) in 
a new fermions and bosons \cite{IJQC}. This leads to new V-E Hamiltonian (we
omit - on the second quantized operators)

\begin{equation} \label{e28}
H \ = \ H_{A} \ + \ H_{B}
\end{equation}
where
\begin{eqnarray} \label{e29}
H_{A} & = & E_{NN}(B) \ - \ E_{NN}^{(2)}(B) \ - \ V_{N}^{(2)} (B) \ + \
\sum_{RSI} h_{RS} (B) C_{RI}C_{SI} \nonumber \\[2mm]
&+&  \frac{1}{2} \ \sum_{RSTUIJ} \ (v_{RTSU}^{0}- v_{RSTU}^{0})
 C_{RI} C_{SI} C_{TJ} C_{UJ} \nonumber \\[2mm]
&+& \sum_{PQRS} h_{RS}C_{RP}C_{SQ} N \left[ a_{P}^{+} a_{Q} \right]
\nonumber \\[2mm]
&+& \sum_{PQRSTUI}(v_{RTSU}^{0}- v_{RSTU}^{0})C_{RP}C_{SQ}C_{TI}C_{UI}
 N \left[ a_{P}^{+} a_{Q} \right] \nonumber \\[2mm]
&+& \sum_{PQRSTUVW} v_{TUVW}^{0} C_{TP}C_{UQ}C_{VR}C_{WS} 
N \left[ a_{P}^{+} a_{Q}^{+} a_{S} a_{R} \right]
\end{eqnarray}
and
\begin{eqnarray} \label{30}
H_{B} & =&  \sum_{r} \ \hbar \omega_{r} \ \left( b_{r}^{+} b_{r} \ + \ \frac{1}{2}
\right)  \nonumber \\[2mm]
&+&  \sum_{PQr} \ \hbar \omega_{r}  \left( b_{r}^{+} d_{rPQ}  + 
d_{rQP} b_{r} \right) \ N \ \left[ a_{P}^{+} a_{Q} \right]   \nonumber \\[2mm]
&+&  \sum_{AIr} \ \hbar \omega_{r} \ \left( d_{rAI} \right)^{2} \ + \
\sum_{PQAIr} \ \hbar \omega_{r} \ \left( d_{rPA} d_{rQA} \ - \
d_{rPI} d_{rQI} \right) \ N \ \left[ a_{P}^{+} a_{Q} \right] \nonumber \\[2mm]
&+&  \sum_{PQRSr} \ \hbar \ \omega_{r} d_{rPS} d_{rQR}  N  \left[
a_{P}^{+} a_{Q}^{+} a_{S} a_{R} \right] 
\end{eqnarray}

If we introduce the following quantities
\begin{eqnarray} \label{e31}
E_{SCF} & =&  \sum_{RSI} h_{RS} C_{RI} C_{SI}   \nonumber \\[2mm]
&+& \ \frac{1}{2} \sum_{RSTUIJ} \left( v_{RTSU}^{0} \ - \ v_{RSTU}^{0} \right)
C_{RI} C_{SI} C_{TJ} C_{UJ} 
\end{eqnarray} 
the new Hartree-Fock operator  $f$  with the matrix elements
\begin{eqnarray} \label{e32}
f_{PQ} & =& \ \sum_{RS} h_{RS} C_{RP} C_{SQ}  \nonumber \\[2mm]
&+&  \sum_{RTSUI} \left( v_{RTSU}^{0} \ - \ v_{RSTU}^{0} \right) C_{RP} C_{SQ}
C_{TI} C_{UI} 
\end{eqnarray}
and the new two-particle integral
\begin{equation} \label{e33}
v_{PQRS} \ = \ \sum_{TUVW} v_{TUVW}^{0} C_{TP} C_{UQ} C_{VR} C_{WS} \ .
\end{equation}

We can rewrite our Hamiltonian \ $H_{A}$ \ (\ref{e29}) to the form
\begin{eqnarray} \label{e34}
H_{A} & =&  E_{NN} \ - \ E_{NN}^{(2)} \ - \ V_{N}^{(2)} \ + \ E_{SCF}  + 
\sum_{PQ} f_{PQ} N \left[ a_{P}^{+} a_{Q} \right]   \nonumber \\[2mm]
&+&  \frac{1}{2} \sum_{PQRS} v_{PQRS}N \left[ a_{P}^{+} a_{Q}^{+} a_{S} a_{R} \right] 
\end{eqnarray}
Here in eq. (\ref{e34})  \( \sum_{PQ} f_{PQ} N \left[ a_{P}^{+} a_{Q} \right] \) 
is new Hartree-Fock operator for a new fermions (\ref{e21}), (\ref{e22}), operator
  \( \frac{1}{2} \sum_{PQRS} v_{PQRS}N \left[ a_{P}^{+} a_{Q}^{+} a_{S} a_{R} 
\right] \)  is a new fermion correlation operator and  $E_{SCF}$  is a new fermion
Hartree-Fock energy. Our new basis set is obtained by diagonalizing the 
operator $f$ from eq. (\ref{e32}). The new Fermi vacuum is renormalized Fermi vacuum
and new fermions are renormalized electrons. The diagonalization
of $f$ operator (\ref{e32}) leads to coupled perturbed Hartree-Fock (CPHF)
equations \cite{cphf,Gerratt1,Gerratt2}. Similarly operators $\stackrel{-}{b}_{r}$  
( $\stackrel{-}{b}_{r}^{+}$ ) corresponds to renormalized phonons.
Using the quasiparticle canonical transformations (\ref{e21} -\ref{e24})
 and the Wick
theorem the V-E Hamiltonian takes the form
\begin{equation} \label{e35}
H \ = \ H_{A} \ + \ B_{B}
\end{equation}
where
 
\begin{eqnarray} \label{e37}
H_{A} & =&  E_{NN}^{0}  +  E_{SCF}^{0}  +  \sum_{P} \epsilon_{P}
N \left[ a_{P}^{+} a_{P} \right]    \nonumber \\[2mm]
&+& \sum_{n=1}
^{\infty} \ \sum_{k=0}^{[n/2]} \ {\bf E}^{(k,n-2k)} \ . \ {\bf B}^{(n-2k)}
  \nonumber \\[2mm]
&+& \sum_{n=1}^{\infty} \
\sum_{k=0}^{[n/2]} \ \sum_{PQ} \ {\bf f}_{PQ}^{(k,n -2k)} \ {\bf B}^{(n -2k)} \
N \left[ a_{P}^{+} a_{Q} \right] \nonumber \\[2mm]
&+& \frac{1}{2} \sum_{n=0}^{\infty} \ \sum_{k=0}^{[n/2]} \ \sum_{PQRS} \
{\bf v}_{PQRS}^{(k,n -2k)} \ . \ {\bf B}^{(n - 2k)} \ N \left[ a_{P}^{+} a_{Q}
^{+} a_{S} a_{R} \right]  . 
\end{eqnarray}
and
\begin{eqnarray} \label{e38}
H_{B} & =&  \sum_{r} \ \hbar \omega_{r} \ \left( b_{r}^{+} b_{r} \ + \ \frac{1}{2}
\right)  \nonumber \\[2mm]
&+&  \sum_{AIr} \ \hbar \omega_{r} \ \left( d_{rAI} \right)^{2} \ + 
  \sum_{PQr} \ \hbar \omega_{r}  \left( b_{r}^{+} d_{rPQ}  +
d_{rQP} b_{r} \right) \ N  \left[ a_{P}^{+} a_{Q} \right]  
 \nonumber \\[2mm]
&+&\sum_{PQAIr} \ \hbar \omega_{r}  \left( d_{rPA} d_{rQA} \ - \
d_{rPI} d_{rQI} \right) \ N  \left[ a_{P}^{+} a_{Q} \right]
 \nonumber \\[2mm]
&+&  \sum_{PQRSr} \ \hbar \ \omega_{r} d_{rPS} d_{rQR}  N  \left[
a_{P}^{+} a_{Q}^{+} a_{S} a_{R} \right] \ .
\end{eqnarray}

As we have shown in \cite{IJQC,thesis} this quasiparticle transformation leads from 
crude adiabatic to adiabatic Hamiltonian. This Hamiltonian (\ref{e35}) is adiabatic 
Hamiltonian. Note that the force constant for harmonic 
oscilators is given as second derivative of $E_{SCF}$  at point $R_{o}$ .
We shall call the corresponding phonons as adiabatic phonons.

\section{Diabatic canonical transformation}

In previous part we developed canonical transformation 
(through normal coordinates) by which we were able to pass from
crude adiabatic to adiabatic Hamiltonian. We started with crude 
adiabatic molecular Hamiltonian on which we applied canonical transformation
on second quantized operators
\begin{equation} \label{e39}
\stackrel{-}{a}_{P}  =  \sum_{Q} C_{PQ} (B) a_{Q}
\end{equation}
\begin{equation} \label{e40}
\stackrel{-}{a}_{P}^{+}  =  \sum_{Q} C_{PQ} (B)^{+} a_{Q}^{+}
\end{equation}
\begin{equation} \label{e41}
\stackrel{-}{b}  =  b_{r} \ +  \sum_{PQ} d_{rPQ} (B) a_{P}^{+} a_{Q}
\end{equation}
\begin{equation} \label{e42}
\stackrel{-}{b}_{r}^{+}  =  b_{r}^{+} \ + \ \sum_{PQ} d_{rQP} (B)^{+}
a_{P}^{+} a_{Q} \ .
\end{equation}
where operators  $\stackrel{-}{a}_{P} (\stackrel{-}{a}_{P}^{+}$) 
corresponds to fermions and operators \ $\stackrel{-}{b} 
(\stackrel{-}{b}_{r}^{+})$  to bosons and  $B=b+b^{+}$  is the normal
 coordinate.
The coefficients   $C_{PQ}$ ($d_{rPQ}$) can be found from the solution
of CPHF equations. We also found that adiabatic corrections can be
calculated as perturbation corrections, which mean that we expect that
 adiabatic corrections  represents small perturbation. The situation can be
 more complex if we cannot treat non-adiabaticity as a perturbation. This is
the case when non-adiabaticity can cause strong coupling between
two or more electronic states. In order to treat such situations we
can procede in a way in which we generalize transformations (\ref{e21})-(\ref{e24}). In
these equations the expansion coefficients $C$ and $d$ were functions
of normal coordinates $B=b+b^{+}$. 

The generalization can be done in
a way that these coefficients are some general functions $C_{PQ} (b,b^{+})$
and $d_{rPQ} (b,b^{+})$ of $b$ and $b^{+}$ operators. We can expect that
these coefficients will be not only the function of normal coordinate
$B=b+b^{+}$ but also the function of momentum $  \tilde{B} =b-b^{+}$.
 Therefore general transformations will have the form \cite{molvib,thesis}
\begin{equation} \label{e43}
\stackrel{-}{a}_{P} \ =  \sum_{Q} C_{PQ} (B,\tilde {B}) a_{Q}
\end{equation}

\begin{equation} \label{e44}
\stackrel{-}{a}_{P}^{+}  =  \sum_{Q} C_{PQ} (B,\tilde {B})^{+} a_{Q}^{+}
\end{equation}

\begin{equation} \label{e45}
\stackrel{-}{b}_  =  b_{r}  +  \sum_{PQ} d_{rPQ} (B,\tilde {B}) a_{P}^{+} a_{Q}
\end{equation}

\begin{equation} \label{e46}
\stackrel{-}{b}_{r}^{+}  =  b_{r}^{+} +  \sum_{PQ} d_{rQP} (B,\tilde {B})^{+}
a_{P}^{+} a_{Q} 
\end{equation}

Such transformations would be rather complex, therefore we try simple 
approximation

\begin{equation} \label{e47}
C (B,\tilde {B})=C (B). \tilde {C} (\tilde {B})
\end{equation}

Further we can procede similarly as in the case of adiabatic approximation.
We shall not present here the details. These are presented in 
\cite{molvib,thesis}. We just 
mention the most important features of our transformation (\ref{e43}-\ref{e47}).
Firstly, when passing from crude adiabatic to adiabatic approximation
the force constant changed from second derivative  of electron--nuclei
interaction  $u_{SCF}^{(2)}$  to second derivative of Hatree--Fock energy
 $E_{SCF}^{(2)}$  . Therefore when performing transformation (\ref{e43}-\ref{e47}) we expect
change of  force constant and therefore change of the vibrational part of 
Hamiltonian

\begin{equation} \label{e48}
       H_{B} = E_{kin}(\tilde{B})
              + E_{pot}(B)
\end{equation}

The potential energy is determined by the quadratic part of the nuclear
energy  $E_{NN}^{2}(B)$ as well as by some potential energy
$V_{N}^{(2)}(B)$ which is a quadratic function of coordinate operators
and has its origin in the interaction of the electrons with the 
vibrating nuclei. Therefore we have

\begin{equation} \label{e49}
  E_{pot}(B) = E_{NN}^{(2)}(B) + V_{N}^{(2)}(B)
\end{equation}
In the case of kinetic energy term this was identical with the kinetic 
energy of the nuclei in the case of adiabatic approximation. In the case
of the breakdown of adiabatic approximation we have to remember the
finite mass of electrons and therefore to introduce  more general
kinetic energy term. Therefore, we add to the kinetic energy of the nuclei
$T_{N}(\tilde{B})$ some other yet unknown term which will be the
quadratic function of momentum operator 

\begin{equation} \label{e50}
        E_{kin}(\tilde{B})
        = T_{N}(\tilde{B})
        + W_{N}^{(2)}(\tilde{B})
\end{equation}

The total vibrational--electronic Hamiltonian
\be \label{e51}
        H = H_{A} + H_{B}
\ee 
will have the form
\be \label{e52}
        H_{B}
        = T_{N}^{(2)}(\tilde{B})
        + W_{N}^{(2)}(\tilde{B})
        + E_{NN}^{(2)}(B)
        + V_{N}^{(2)}(B)
\ee
and
\bea \label{e53}
	H_{A} &=& E_{NN}(B)- E_{NN}^{(2)}(B)- V_{N}^{(2)}(B)-W_{N}^{(2)}
	(\tilde{B})
\nonumber
\\[2mm] 
	&+& \sum_{P,Q} h_{PQ}(B)a_{p}^{+} a_{q}
        + \frac{1}{2}\sum_{PQRS} V^{0}_{PQRS}a_{P}^{+}a_{Q}^{+}a_{S}a_{R} 
        \label{eq3}
\eea
Secondly, coefficients $C_{PQ}$ and $\tilde {C}_{PQ}$ 
are determined through equations \cite{molvib} 
\[
        u_{PQ}^{r} + (\varepsilon_{P}^{0}
        - \varepsilon_{Q}^{0})C_{PQ}^{r}
        + \sum_{AI} \,
        \left[ (V_{PIQA}^{0} - V_{PIAQ}^{0})C_{AI}^{r}
        - (V_{PAQI}^{0}-V_{PAIQ}^{0})C_{IA}^{r} \right]
\]
\be \label{e54}
        - \hbar\omega_{r} \tilde{C}_{PQ}^{r}
        = \varepsilon_{P}^{r}\delta_{PQ}
\ee
\[
        (\varepsilon_{P}^{0}
        - \varepsilon_{Q}^{0})\tilde{C}_{PQ}^{r}
        + \sum_{AI} \,
        \left[ (V_{PIQA}^{0}-V_{PIAQ}^{0})\tilde{C}_{AI}^{r}
        - (V_{PAQI}^{0}-V_{PAIQ}^{0})\tilde{C}_{IA}^{r} \right]
\]
\be \label{e55}
        - \hbar\omega_{r} \tilde{C}_{PQ}^{r}
        = \tilde{\varepsilon}_{P}^{r} \delta_{PQ}
\ee

where $\hbar\omega_{r}$ is the new non-adiabatic phonon given by
\be \label{e56}
        H_{B} =
        \sum_{r} \hbar\omega_{r}(b_{r}^{+}b_{r}
        + \frac{1}{2})
        \label{eq4}
\ee

The expressions for extra terms  $V_{N}^{(2)}(B)$
and $W_{N}^{(2)}(\tilde{B})$ in (\ref{e52}) are given as
 
\be \label{e57}
        V_{N}^{rs} = \sum_{I} u_{II}^{rs} + \sum_{IA}
        \left[ (u_{AI}^{r}
        + \hbar\omega_{r}\tilde{C}_{AI}^{r})C_{IA}^{s}
        + (u_{AI}^{r} + \hbar\omega_{s}
        \tilde{C}_{AI}^{r})C_{IA}^{r} \right]
\ee
and
\be \label{58}
        W_{N}^{rs} = 2 \hbar\omega_{r}
        \sum_{AI} C_{AI}^{r}\tilde{C}_{IA}^{s}
\ee
This means that the resulting vibrational frequency $\omega$
depends explicitly on coefficients $\tilde{C}_{PQ}^{r}$ and
$C_{PQ}^{r}$.

Finally fermion part of Hamiltonian will be given as
\be \label{e59}
	H_{F}=H_{F}^{0} + H_{F^{'}} + H_{F^{''}} + H_{F^{'''}}
\ee
For the ground state energy we get
\be \label{e60}
	H_{F}^{0} = E_{NN}^{0} + E_{SCF}^{0} 
	+ \sum_{AIr} \hbar\omega_{r}
	(|C_{A}^{r}|^{2} - |\tilde{C}_{AI}^{r}|^{2})
	\label{sel.eqs.b}
\ee
	One-fermion part will be
\begin{eqnarray} \label{e61}
	H_{F'}&=&\sum_{P} \varepsilon_{P}^{0} 
	N[a_{P}^{+}a_{P}]
	+ \sum_{PQr}\hbar\omega_{r}
	\left[\sum_{A}(C_{PA}^{r}C_{QA}^{r*}
	- \tilde{C}_{PA}^{r}\tilde{C}_{QA}^{r*}) - \right.
\nonumber\\[2mm]
	&-& \left. \sum_{I}
	(C_{PI}^{r}C_{QI}^{r*}-\tilde{C}_{PI}^{r}\tilde{C}_{QI}^{r*})
	\right]
	N[a_{P}^{+}a_{Q}] - 2\sum_{PQr}E^{r*}\tilde{C}_{PQ}^{r}
	N[a_{P}^{+}a_{Q}] 
\nonumber\\[2mm]
	&+& \sum_{PQr} \left[ (h(P)-p(P))\varepsilon_{P}^{r*} 
	+ (h(Q) - p(Q))\varepsilon_{Q}^{r*} \right]
	\tilde{C}_{PQ}^{r}N[a_{P}^{+}a_{Q}]
\nonumber\\[2mm]
	&-& \sum_{PQAIr} \left[ (v_{PIQA}^{r} 
	- v_{PIAQ}^{r})\tilde{C}_{IA}^{r*} 
	+ (v_{PAQI}^{r} - v_{PAIQ}^{r})\tilde{C}_{AI}^{r*}
	\right] N[a_{P}^{+}a_{Q}]
\end{eqnarray}
Two-fermion part will be
\begin{eqnarray} \label{e62}
	H_{F''} &=& \frac{1}{2}\sum_{PQRS} v_{PQRS}^{0}
	N[a_{P}^{+}a_{Q}^{+}a_{S}a_{R}]
\nonumber\\[2mm]
	&+& \sum_{PQRSr}\hbar\omega_{r}(C_{PR}^{r}C_{SQ}^{r*} 
	- \tilde{C}_{PR}^{r}\tilde{C}_{SQ}^{r*}) 
	N[a_{P}^{+}a_{Q}^{+}a_{S}a_{R}]
\nonumber\\[2mm]
	&-& 2\sum_{PQRSr} \varepsilon_{P}^{r}\tilde{C}_{SQ}^{r*}
	N[a_{P}^{+}a_{Q}^{+} a_{S}a_{R}]
\nonumber\\[2mm]
	&+& 2\sum_{PQRSTr} \left\{ 
	\sum_{I} \left[ v_{PQTS}^{0}C_{TI}^{r}
	- v_{PQTI}^{0}C_{TS}^{r} 
	+ (v_{TQSI}^{0}-v_{TQIS}^{0})C_{PT}^{r}
	\right] \tilde{C}_{RI}^{r*} \right. 
\nonumber\\[2mm]
	&+& \sum_{I} \left[ v_{TIRS}^{0}C_{QT}^{r} 
	- v_{TQRS}^{0}C_{IT}^{r}
	+ (v_{IQTS}^{0} - v_{IQST}^{0})C_{TR}^{r}
	\right] \tilde{C}_{IP}^{r*}
\\
	&-& \sum_{A} \left[ v_{PQTS}^{0}C_{TA}^{r} 
	- v_{PQTA}C_{TS}^{r}
	+ (v_{TQSA}^{0}-v_{TQAS}^{0})C_{PT}^{r}
	\right] \tilde{C}_{RA}^{r*} 
\nonumber\\[2mm]
	&-& \left. \sum_{A} \left[ v_{TARS}^{0}C_{QT}^{r} 
	- v_{TQRS}C_{AT}^{r}
	+ (v_{AQTS}^{0}-v_{AQST}^{0})C_{TR}^{r}
	\right] \tilde{C}_{AP}^{r*} \right\}
	N[a_{P}^{+}a_{Q}^{+} a_{S}a_{R}]
\nonumber
\end{eqnarray}
Three-fermion part will be (as a result of transformation (\ref{e47}) the
three fermion term appears)
\be \label{e63}
	H_{F'''} =
	- 2\sum_{PQRSTUVr}(v_{PQVT}^{0}C_{RS}^{r} 
	- v_{VQST}^{0}C_{PV}^{r})
	\tilde{C}_{UR}^{r*} 
	N[a_{P}^{+}a_{Q}^{+}a_{R}^{+}a_{U}a_{T}a_{S}]
\ee
The bosonic part of  Hamiltonian $H_{B}$ is not given in a diagonal form. 
To bring it to diagonal form as in eq. (\ref{e56}) we can proceede as follows.
\begin{eqnarray} \label{e64}
	H_{B} &=& T_{N} + E_{NN}^{(2)}
\nonumber \\[2mm]
	&+& \sum_{r,s}  \stackrel{\wedge}{E}_{SCF}^{rs}  -  \sum_{I} 
\left[
\epsilon_{I}^{0}  \stackrel{\wedge}{S}_{II}^{rs}  +  \frac{1}{2} \ \left(
\epsilon_{I}^{r} \ \stackrel{\wedge}{S}_{II}^{s} \ + \ \epsilon_{I}^{s} \
\stackrel{\wedge}{S}_{II}^{r} \right)  \right] 
\nonumber \\[2mm]
	&+& \sum_{RI}  \left[ \left( \stackrel{\wedge}{f}_{RI}^{r} 
	-  \epsilon
_{I}^{0} \ \stackrel{\wedge}{S}_{RI}^{r} \ + \ \hbar \ \omega_{r} \
\stackrel{\sim}{C}_{RI}^{r} \right)  \right. \stackrel{\wedge}{C}_{RI}^{s} 
\nonumber \\[2mm]
	&+&  \left( \stackrel{\wedge}{f}_{RI}^{s}  -  \epsilon_{I}^{0} 
\stackrel{\wedge}{S}_{RI}^{s} + \hbar \omega_{s} 
\stackrel{\sim}{C}_{RI}^{s} \right)  \stackrel{\wedge}{C}_{RI}^{r}
  B_{r}B_{s} 
\nonumber \\[2mm]
	&+&  \sum_{r,s} 2 \hbar  \omega_{r}  \sum_{AI}  \left(
\stackrel{\wedge}{C}_{AI}^{r} \ + \ <  A  |  I^{r}  >  \right) 
\tilde{C}_{AI}^{s}  \tilde{B}_{r} \tilde{B}_{s}
\end{eqnarray}

Our aim is to bring this Hamiltonian into diagonal form. We can extract 
adiabatic part ($\hbar  \omega_{r}^{a}$) and we get
\be \label{e65}
        H_{B} = \sum_{r} \, \hbar\omega_{r}^{a}
        ( b_{r}^{+}b_{s} + \frac{1}{2} )\delta_{rs}+ \ F_{rs}^{1} + \ F_{rs}^{2}
\ee
where
\begin{equation} \label{e66}
F_{rs}^{1}= 2 \sum_{r,s} \left \{\sum_{R,I}
\left [(\stackrel{\wedge}{f}_{RI}^{s} \ - \ \epsilon_{I}^{0} \ )(\stackrel{\wedge}{C}_{RI}^{s}-\stackrel{\wedge}{\bar{C}}_{RI}^{r})+ 
\hbar \ \omega_{r}\tilde {C}_{RI}^{r}\tilde {C}_{RI}^{s} \right ] \right \}B_{r}
B_{s}
\end{equation}

\begin{equation} \label{e67}
F_{rs}^{2}= 2 \sum_{r,s}\hbar \ \omega_{r} \left \{\sum_{AI} \ (
\stackrel{\wedge}{C}_{AI}^{r} \ + \ < \ A \ | \ I^{r} \ > \ ) \
\stackrel{\wedge}{C}_{AI}^{s} 
\right \} \tilde {B}_{r}
\tilde {B}_{s}
\end{equation}
where $\stackrel{\wedge}{\bar {C}}_{RI}^{r}$ is identical with 
$\stackrel{\wedge}{C}_{RI}^{r}$
coefficients from adiabatic transformation eqs. (\ref{e19}, \ref{e20}).
 If we substitute in eq. (\ref{e66})
 and (\ref{e67}) for
$B_{r}=b_{r}^{+} +b_{r}$ and $\tilde {B}_{r}=b_{r}-b_{r}^{+}$ we get for
(\ref{e65}) the expression 
\be \label{e68}
	H_{B} = \sum_{rs} \,
	\left[ A_{rs}b_{r}^{+}b_{s} + \frac{1}{2}B_{rs}
	(b_{r}^{+}b_{s}^{+} + b_{r}b_{s}) \right]
\ee
where 
\be \label{e69}
	B_{rs} = 2 (F_{rs}^{1} + F_{rs}^{2})
\ee
\be \label{e70}
	A_{rs} = F_{rs}^{1} + F_{sr}^{1} - F_{rs}^{2} - F_{sr}^{2}
\ee
Diagonalizing the above Hamiltonian we obtain diabatic frequencies
\be \label{e71}
	H_{B} = \sum_{r} \, \hbar\omega_{r}^{d} 
	(b_{r}^{+}b_{r} + \frac{1}{2} )
\ee
Hamiltonian (\ref{e68}) has a form of quadratic Hamiltonian 
\cite{Blaizot86,Asian}
and can be diagonalized by Bogoljubov transformation, 
 which leads to the condition
 
\be \label{e72}
        det \left( \begin{array}{cc}
        \hat{A}-\omega^{d}\hat{1} & \hat{B} \\
        -\hat{B} & -\hat{A}-\omega^{d} \hat{1}
        \end{array} \right) =0
\ee

Secular equation (\ref{e72}) gives us diabatic phonons $\hbar\omega_{r}^{d}$. \\
If we look at eq. (\ref{e38}) we see that we have corrections due to non--adiabaticity
to one-particle part as well as to two-particle part. 
 We see the hierarchical structure of our Hamiltonian.
If the non--adiabatic coupling is small i.e. $\tilde{C}$ goes to zero
and we have adiabatic Hamiltonian. If this coupling is strong we cannot use 
adiabatic approximation but we have to work with full V-E Hamiltonian 
(\ref{e51}).

\section{Calculations}

In order to compare our approach with other approaches dealing with
adiabatic corrections we perform simple 
model calculations for adiabatic corrections to ground state energy.
We start with adiabatic Hamiltonian (\ref{e28}). We now perform the following approximation.
We limit ourselves to finite orders of Taylor expansion of the operators
 $H_{A}^{'}$  and  $H_{B}^{'}$ . We shall use similar approximation 
as in \cite{JER}. 
The diagrammatic representation of our approximate Hamiltonian will be
\begin{eqnarray} \label{e73}
H=H_{o}+H^{'} & =&  E_{NN}^{o}\ +  E_{SCF}^{o}  
\sum_{P} \epsilon_{P} N \left[ a_{P}^{+} a_{P} \right]  \nonumber \\[2mm]
&+&  \sum_{r} \ \hbar \omega_{r}  \left( b_{r}^{+} b_{r}  +  \frac{1}{2}
\right)    
\end{eqnarray}
\\[40mm]
{\it ---------------------Here should be included image.gif-----------------------}
\\[40mm]

The adiabatic corrections to the ground state of $H_{2}$, $HD$, and $D_{2}$ we shall
calculate using second--order Rayleigh--Schr\"{o}dinger many-body
 perturbation theory (RS--MBPT)
and our Hamiltonian (\ref{e73}).
If we assume that we know the solution of the unperturbed Schr\"odinger equation
\begin{equation} \label{e74}
H_{0} \mid \varphi _{o}  > \ = \ E_{o} \mid \varphi _{o}  > \ ,
\end{equation}
where \ $H_{0}$ \ is the unperturbed Hamiltonian 
\ $H_{A}^{o}$ \ + \ $H_{B}^{o}$ \
where
\begin{equation} \label{e75}
H_{A}^{o} \ = \ E_{NN}^{o}\ + \ E_{SCF}^{o} \
+ \sum_{P} \epsilon_{P} N \left[ a_{P}^{+} a_{P} \right] 
\end{equation}
and
\begin{equation} \label{e76}
H_{B}^{o} \ = \sum_{r} \ \hbar \omega_{r}  \left( b_{r}^{+} b_{r}  +  \frac{1}{2}
\right)    
\end{equation} 
The perturbed (exact) Schr\"odinger equation will read
\begin{equation} \label{e77}
H \mid \Psi  > \ = {\cal E } \mid \Psi  > \ ,
\end{equation}
where $H$ will be our Hamiltonian (\ref{e73}). The perturbed energy $E$ will
be given through the RS--MBPT expansion as

\begin{equation} \label{e78}
{\cal E } = \ E_{o} + < \varphi _{o} \mid \ H^{'} \mid \varphi _{o} \ > \ + \
< \ \varphi_{o} \mid \ H^{'} Q_{o}  H^{'} \mid \varphi_{o} \ > \ + \ \ldots \ ,
\end{equation}
where  $H^{'}$  is the perturbation and   $Q_{o}$  is the
resolvent
\begin{equation} \label{e79}
Q_{o} \ = \ \sum_{i \neq 0} \frac{ \mid \varphi _{i} > < \varphi _{i} \mid }
{E_{0} - E_{i}}
\end{equation}
Since our sets of boson creation and annihilation operators and fermion creation
and annihilation operators commute we can write our unperturbed wavefuntion
\ $\mid \varphi _{o} >$ \ as the product of the fermion state vector  $\mid
\psi _{o} >$  and the boson state vector  $\mid \chi _{o} >$  , i.e.
\begin{equation} \label{e80}
\mid \varphi _{o}  >  =  \mid \psi _{o}  > \ \mid \chi _{o} >
\end{equation}

Further we want to study the nonadiabatic corrections to the ground state.
Therefore   $\mid \psi _{o} >$  will
be the unperturbed ground state wave function (we shall use Hartree--Fock 
ground state Slater determinant --Fermi vacuum) and  $\mid \chi _{o}  >$  
will be boson ground state --boson vacuum  $\mid 0  >$.

\begin{equation} \label{e81}
\ \mid \chi _{0}  > = \ \mid   0  >  .
\end{equation}

The exact ground state energy will be given by perturbation expansion
(up to the second order)
\begin{eqnarray} \label{e82}
{\cal E } &=& < \psi_{0} \mid <  0 \mid \ H_{0} \mid 0  > \
\mid \ \psi_{0}  >  \nonumber \\[2mm]
&+& \ <  \psi_{0} \mid \ <  0 \mid \ H^{'}  
\mid \ 0  > \mid \ \psi_{0}  >  \nonumber \\[2mm]
&+& \ <  \psi_{0} \mid \ <  0 \mid \ H^{'}  Q_{0} H^{'}
\mid \ 0  > \mid \ \psi_{0}  > + \ldots 
\end{eqnarray}

Substituting for  $H_{o}=H_{A}^{o} + H_{B}^{o}$ from (\ref{e75}, \ref{e76})
 into the first term in (\ref{e82}) we get

\begin{equation} \label{e83}
< \psi_{0} \mid <  0 \mid \ H_{o} \mid 0 \ > \mid \psi_{0} \ > \
=  \ E_{NN}^{o}\ + \ E_{SCF}^{o} \ + \sum_{r} \ \frac{1}{2} \hbar \omega_{r}
\end{equation}

We shall not present all terms for perturbation corrections from the
right hand side of eq. (\ref{e82}). There are corrections which
 corresponds to electron correlation,
anharmonicity corrections and adiabatic corrections \cite{JER}. We shall pay
attention only to adiabatic corrections given through second-order
term in eq. (\ref{e82}). 
Analyzing diagrammatic contributions through the Hamiltonian (\ref{e73}) we find 
that the adiabatic corrections are given through the 
second up to fourth term in eq. (\ref{e38}). From these terms we calculate
only contributions from the first and the second term which are given 
through the first order of Taylor expansion and these terms are used in 
second order RS--MBPT.
We obtain the following simple expressions

\begin{eqnarray} \label{e84}
< \psi_{0} \mid <  0 \mid  H^{'} \mid 0  > \mid \psi_{0}  >  & \sim &
 \nonumber\\[2mm]
&& \sum_{AIr} \hbar \omega_{r} \left( \stackrel{\wedge}{C}_{AI}^{r} + <
A (0) \mid \stackrel{\wedge}{I}^{r}  >
\right)^{2} \nonumber \\[2mm]  
&+&  \sum_{AIr} \hbar^{2} \omega_{r}^{2} \left( \stackrel{\wedge}{C}_{AI}^{r} + <
A (0) \mid \stackrel{\wedge}{I}^{r}  >
\right)^{2} \ . \ \left(\varepsilon_{I}-\varepsilon_{A}- \hbar \omega_{r} 
\right)^{-1} \nonumber \\[2mm] 
&+&  \sum_{AIr} \hbar^{2} \omega_{r}^{2} \left( \stackrel{\wedge}{C}_{AI}^{r} + <
A (0) \mid \stackrel{\wedge}{I}^{r}  >
\right)^{2} \ . \ \left(\varepsilon_{I}-\varepsilon_{A}+ \hbar \omega_{r}
 \right)^{-1} \nonumber \\[2mm] 
&\approx & \sum_{AIr} \hbar \omega_{r} \left( \stackrel{\wedge}{C}_{AI}^{r} + <
A (0) \mid \stackrel{\wedge}{I}^{r}  >
\right)^{2} \nonumber \\[2mm]  
&+& 2 \sum_{AIr} \hbar^{2} \omega_{r}^{2} \left( \stackrel{\wedge}{C}_{AI}^{r} + <
A (0) \mid \stackrel{\wedge}{I}^{r}  >
\right)^{2} \ . \ \left(\varepsilon_{I}-\varepsilon_{A} \right)^{-1} \ 
\end{eqnarray}
For the notation see \cite{molvib,JER}. We believe these three terms on right
hand side of (\ref{e84}) represents the dominant contributions to adiabatic
corrections. The last formula is valid due to the eq. (\ref{BO}).
 This formula was used to calculate the adiabatic
corrections to the ground state energy of the $H_{2}$,  $D_{2}$,
and  $HD$, molecules. Results in $cm^{-1}$ are presented in Table I.
We see  that the structure of this formula is similar to
eq. (\ref{B})  except that we use RS--MBPT. We also used
the same approach to calculate the adiabatic corrections to the energies
of the first vibrational transitions for the same molecules \cite{JER}. Analyzing 
eq. (\ref{e84}) we can see
that the first term on the right hand side of eq.(\ref{e84}) is always
positive and correspond to largest contribution, while
the second term on right hand side of eq. (\ref{e84}) is always negative
and represents smaller contribution than the previous term.
Therefore we can expect that the eq. (\ref{e84}) should converge to the true
value of adiabatic correction from above. This also explain the
larger values for $H_{2}$,  $D_{2}$, and $HD$ adiabatic correction
obtained through eq. (\ref{e84}) than true value obtained by
 Wolniewicz \cite{Woln}.  Another source of difference can be basis set used
and also the contribution from other terms in (\ref{e38}). 
In our calculation we have obtained for $H_{2}$ value of 136.89 $cm^{-1}$ 
using Roos Augmented Triple Zeta ANO 
\cite{Roos} basis set. Handy \cite{Handy} using basis set of similar
quality as used in our calculations obtained value 101 $cm^{-1}$.
Similar value of adiabatic correction was obtained also in an older
study by Wolfsberg \cite{wolf}.
Kutzelnigg \cite{cenkutz} in his paper using wave function expansion consisting 1200 
functions obtained Wolniewicz \cite{Woln} value 114.591 $cm^{-1}$. It is apparent that
calculations of adiabatic corrections are strongly basis set
dependent.

\section{Conclusions}

In this article we performed simple model calculations of adiabatic 
corrections for ground state energy of the $H_{2}$,  $HD$, and $D_{2}$
molecules. The corrections were derived through canonical transformation
applied to crude adiabatic molecular Hamiltonian. These transformations mix
together electrons and phonons (normal coordinate canonical transformation)
leading to adiabatic molecular Hamiltonian. Using second quantization
formalism and many--body diagramatic perturbation theory and splitting
the adiabatic Hamiltonian into unperturbed part and perturbation we
derived the formulae for adiabatic corrections. The results were compared
with the obtained by different approaches by Wolniewicz \cite{Woln}
and recently by Kutzelnigg \cite{cenkutz}. The quasiparticle canonical
transformations were then generalized in a way that electrons and phonons
are mixed not only through the normal coordinate but also through
the momenta. This canonical transformation leads to non--adiabatic
molecular Hamiltonian (motion of electrons does not follow the motion of
 nuclei, the electrons are phase shifted with respect to nuclei). One can
clearly see that the electronic and vibrational motion 
cannot be separated. The mixed system behaves as one whole
quasiparticle (mixed electrons and phonons through the last
(momentum) transformation behaves as a renormalized
fermions and a different mixture of electrons and phonons
leads to renormalized bosons). This in some extent analogous to the
introduction of quasiparticles in the solid state theory, where the
"bare" electron interacting with quantized lattice vibrations is 
renormalized to "absorb" some part of this interaction, and this quasiparticle
 is known as a polaron. 
We were able to derive equations
for non--adiabatic $\tilde{C}$ coefficients, which permits
us to calculate the so called mass polarization terms and thus
non--adiabatic phonons. It is interesting that the quasiparticles
preserves some interesting features known from pure electronic
molecular Hamiltonian calculations e.g. we can speak about
orbital energies of a new quasipartcles, correlation energies of
 of a new quasipartcles, Hartree--Fock energy (holes and particles),
etc. Further very important property which follow from the last
canonical (momentum) transformation is that we clearly see that
in the case of electronic quasidegeneracy when $\tilde{C}$ coefficients
are non-negligible (non--adiabaticity is not a small correction)
we should work with a full non--adiabatic Hamiltonian. In the
case when non--adiabaticity is a small correction and $\tilde{C}$ 
coefficients are negligible, we can work with adiabatic Hamiltonian
(we have only $C$ coefficients through CPHF equations) and only
if system is perfectly separable that even $C$ coefficients  are
negligible we can work with purely electronic Hamiltonian.

\section*{Acknowledgments}

This work was supported by the grants 1/4197/97 of
the Slovak Grant Agency for Science
and 202/98/1028 of the Grant Agency of the Czech Republic.

\newpage
\begin{table}
\caption{Adiabatic corrections (in $cm^{-1}$) for $H_{2}$, $D_{2}$ and $HD$
molecules.}
\begin{tabular}{||c|c|c|c|c|c||} 
--&Wolfsberg \cite{wolf}&Wolniewicz \cite{Woln} & Kutzelnigg \cite{cenkutz} & 
 Handy \cite{Handy} & Present method, eq.(\ref{e84}) \\ \hline
 $H_{2}$& 101.3 & 114.591  & 114.591 & 101  & 136.89  \\ \hline
$D_{2}$ & 50.7 & 57.296$^{*}$ & 57.296$^{*}$ & 50.5$^{*}$ & 68.62  \\ \hline
$HD$ & 76.0 & 85.943$^{*}$ & 85.943$^{*}$ & 75.8 $^{*}$ & 102.67$^{*}$ \\ 
\end{tabular}
\end{table}
$^{*}$These adiabatic corrections were obtained by reduced mass rescaling from
$H_{2}$ corrections \cite{wolf}.

\end{document}